\documentstyle[aps,eqsecnum,amsfonts,preprint]{revtex}
\begin{document}
\title{ \hfill OKHEP-95-07\\
Quasi-Local Formulation of Non-Abelian Finite-Element Gauge Theory}
\author{Kimball A. Milton\thanks{E-mail: kmilton@uoknor.edu}}
\address{\it Department of Physics and Astronomy,
The University of Oklahoma, Norman OK 73019}

\date{\today}
\maketitle

\begin{abstract}
Recently it was shown how to formulate the finite-element
equations of motion of a non-Abelian gauge theory, by
gauging the free lattice difference equations, and simultaneously
determining the form of the gauge transformations.  In
particular, the gauge-covariant field strength was explicitly
constructed, locally, in terms of a path ordered product
of exponentials (link operators).  On the other hand, the
Dirac and Yang-Mills equations were nonlocal, involving sums
over the entire prior lattice.  Earlier, Matsuyama had proposed
a local Dirac equation constructed from just the above-mentioned
link operators.  Here, we show how his scheme, which is closely
related to our earlier one, can be implemented for a non-Abelian
gauge theory. Although both Dirac and Yang-Mills equations are
now local, the field strength is not.
 The technique is illustrated with a direct
calculation of the current anomalies in two and four space-time
dimensions.  Unfortunately, unlike the original finite-element
proposal, this scheme is in general nonunitary.
\end{abstract}

\section{Introduction}
An alternative approach to lattice field theories, based on the
finite-element equations of motion, has been under development
for over a decade.  (For a recent review see \cite{review}.)
Shortly after the introduction of this method, it was seen how
a Abelian gauge field could be coupled to a fermion in this way
\cite{qed}. The resulting Dirac equation was nonlocal:
In Minkowski space-time, the
term proportional to $\gamma^j$ involved a sum over all values
of the corresponding lattice coordinate, $m_j$, $1\le m_j \le M$,
where $M$ is the number of lattice sites in the $j$ direction,
while the term proportional to $\gamma^0$ involved a sum over
all previous times, $0\le n'\le n$, where $n$ is the current lattice
time.  Shortly after our paper appeared, Matsuyama \cite{mat}
proposed a local finite-element Dirac equation for QED, based
on the immediate introduction of link operators into the free
Dirac equation.  Although it could be argued that this latter
approach was somewhat unnatural because it introduced interactions
into the mass term, the primary reason this idea was not pursued
was that it was quite unclear what the form of the non-Abelian
gauge transformations should be on the finite-element lattice.

Instead, the first foray into non-Abelian finite element
gauge theory \cite{nagt} was based on straightforward gauging
of the global phase symmetry of the free finite-element
Dirac equation.  The form of the interacting Dirac equation
was determined, nonlocally as in the Abelian case, and at the
same time, the form of the gauge transformations of the vector
and scalar potentials was determined, in terms of an infinite
sequence of nested commutators.  The Yang-Mills equations
were determined analogously.  The only thing not explicitly
determined at the time was the form of the construction of
the field strength tensor in terms of the potentials; although
it was perfectly clear that the process could be continued
indefinitely, only the first four terms in the sequence in
powers of potentials were given.  Although \cite{nagt} had
been restricted to (1+1) dimensions for simplicity, that restriction
was easily removed \cite{sing}.

The completion of this construction was only given this spring
\cite{nagt2}.  The essential element was the recognition that
under the previously-determined gauge transformations, a suitable
link operator transformed appropriately.  These operators
then can be used to transform the field strength $F_{\mu\nu}$
averaged over the finite-element hypercube
to the $\mu$, $\nu$ plane where it may be expressed as a path-ordered
product of link operators around a plaquette.

This development makes it possible to revisit Matsuyama's scheme
\cite{mat}.  We will see that it is not only possible to formulate
a local Dirac equation in the non-Abelian regime, but
local gauge-covariant Yang-Mills equations as well.  The resulting
equations are inequivalent to those given previously \cite{nagt2},
but not so different either, for the previous equations were
quasilocal, as seen in the simple difference equation given
for the interaction terms.  But the new formulation is still
{\it nonlocal\/} in that the field strength that appears in the
Yang-Mills
equation involves the vector potential over the entire previous
lattice.

In the next section we restate the gauge transformation properties
of the link operators, and give the corresponding construction
of the field strength.  Then, in Sec.~3 we restate Matsuyama's
prescription for the Dirac equation, followed by the corresponding
local Yang-Mills equation. The resulting nonlocality in the
construction
of the field strength is shown.
A simple
calculation of the axial-vector anomaly in two  dimensions is given
in
Sec.~4.  However, it is not clear how to extend such calculations
to four dimensions because, unlike the original formulation
\cite{mmsb},
in general interactions here break unitarity.   In a particular gauge
in which the transfer matrix is unitary, the current anomalies are
computed in the smallest nontrivial four-dimensional lattice in
Sec.~5.
The corresponding calculation in the standard finite-element
formulation
is given in Sec.~6.
A discussion of how symmetry breaking occurs here is given in the
Conclusion and the Appendix.

\section{Gauge-covariant Link Operators and Construction
of the Field Strength}
It is convenient to define the link operator using coordinates
referring to a given finite element:
\begin{equation}
(L_\mu)_{ijkl}=e^{-igh({\cal A}_\mu)_{m_1+i,m_2+j,m_3+k,m_4+l}},
\label{link}
\end{equation}
where $h$ is the lattice constant, and the vector potential
is an appropriately averaged one:
\begin{equation}
{\cal A}^\mu_{m_\mu,\text{m}_\perp}
={1\over2}\left(A^\mu_{m_\mu,\text{m}_\perp}
+A^\mu_{m_\mu-1,\text{m}_\perp}\right).
\end{equation}
The notation here is that $\text{m}_\perp$ refers to space-time
lattice coordinates other that the one singled out, $m_\mu$.
Each index in (\ref{link}) takes on two values, 0 or 1.
The result of a detailed, constructive calculation
\cite{nagt2} is the
following simple transformation law for the link operator
\begin{equation}
\delta(L_\mu)_{1000}=ig[\delta\omega_{0000}(L_\mu)_{1000}
-(L_\mu)_{1000}\delta\omega_{1000}],
\label{linktrans}
\end{equation} where we have assumed that the first coordinate
index refers to the $\mu$ direction.
Then, it is easy to see that the ``transversely-local''
field strength ${\cal F}_{\mu\nu}$ is given by the following
path-ordered product of link operators around the $\mu$, $\nu$
``plaquette'':
\begin{equation}
e^{-igh^2({\cal F}_{\mu\nu})_{\text{m}}}
=Pe^{-ig\oint A\cdot
dl}=(L_\mu)_{1000}(L_\nu)_{1100}(L_\mu^\dagger)_{1100}
(L_\nu^\dagger)_{0100},
\end{equation}
where the first index is the $\mu$ coordinate and the second index
the
$\nu$ coordinate.
To construct the full field strength, which is forward-averaged
over a hypercube with lower left-hand corner at $\text{m}$
we use the gauge covariant average operators constructed from
the links:
\begin{equation}
(\tilde D_\lambda {\cal F}^{\mu\nu})_{0000}={1\over2}\left[
(L_\lambda)_1
({\cal F}^{\mu\nu})_1
(L^\dagger_\lambda)_1
+({\cal F}^{\mu\nu})_0\right],
\label{covave}
\end{equation}
where on the right side we have only displayed the $\lambda$
coordinate
(the rest are 0).
That is, with an overbar representing forward averaging,
\begin{equation}
x_{\overline m}={1\over2}(x_m+x_{m+1}),
\label{ave}
\end{equation}
the field strength is given by\footnote{The averaging over the finite
element on the left side of (\ref{fs}), required by the
finite-element
prescription, is necessary to ensure unitarity.  If
$(F_{\mu\nu})_{\overline{\text{m}}}$ were replaced by
$(F_{\mu\nu})_{\text{m}}$, even the free theory
would not be unitary.}
\begin{equation}
(F_{\mu\nu})_{\overline{m}}
=\left(\prod_{\lambda\ne\mu,\nu}
\tilde D_\lambda{\cal F}_{\mu\nu}\right)_{\text{m}},
\label{fs}
\end{equation}
 where symmetrical averaging is to be understood.
It is immediately obvious that $F_{\mu\nu}$ given by
(\ref{fs}) transforms covariantly,
\begin{equation}
(\delta F^{\mu\nu})_{\overline{\text{m}}}=ig[\delta\omega_{\text{m}},
(F^{\mu\nu})_{\overline{\text{m}}}].
\label{fsxfr}
\end{equation}

\section{Local Formulation of Yang-Mills Equations}
Matsuyama had proposed a local finite-element formulation
of a fermion interacting with an Abelian gauge field \cite{mat}.
He began by adopting a local form for the fermionic gauge
transformation,
\begin{equation}
\delta\psi_{\text{m}}=ig\delta\omega_{{\text{m}}}
\psi_{\text{m}},
\end{equation} defined on fields at the lattice sites,
rather than in the middle of the finite element as in
\cite{qed,nagt,nagt2}.  Then covariant derivative and averaging
operators can be defined in terms of the link operators
defined in (\ref{link}):
\begin{mathletters}
\begin{eqnarray}
(D_\mu\psi)_{0000}&=&{1\over h}\left[
(L_\mu)_1\psi_1-\psi_0\right],\\
(\tilde D_\mu\psi)_{0000}&=&{1\over 2}\left[
(L_\mu)_1\psi_1+\psi_0\right].
\end{eqnarray}
\end{mathletters}
Up to an ordering ambiguity then, the gauge covariant Dirac
equation is
\begin{equation}
i\gamma^\mu\prod_{\nu\ne\mu}\tilde D_\nu D_\mu\psi
+\mu\prod_\nu\tilde D_\nu \psi=0.
\label{localdir}
\end{equation}
By virtue of (\ref{linktrans}) this equation (\ref{localdir})
is covariant not only under Abelian gauge transformations, but
under non-Abelian ones as well.  Finally, we can transform
to a gauge-covariant average Dirac field by averaging
operators,
\begin{equation}
\Psi_{\overline{\text{m}}}=\left(\prod_\lambda\tilde D_\lambda
\psi\right)_{\text{m}},
\end{equation}
[$\ne\psi_{\overline{m}}$  using the notation of (\ref{ave})]
which transforms covariantly in the sense of \cite{nagt2},
\begin{equation}
\delta\Psi_{\overline{\text{m}}}=ig\delta\omega_{{\text{m}}}
\Psi_{\overline{\text{m}}}.
\end{equation}
However, in \cite{nagt2} it the ``free'' average of the Dirac
field, $\psi_{\overline{m}}$ that transforms covariantly.

We proceed similarly in the gauge sector.  If the field strength
like the Dirac field transformed locally at the lattice sites,
\begin{equation}
(\delta F^{\mu\nu})_{\text{m}}=ig[\delta\omega_{\text{m}},
(F^{\mu\nu})_{\text{m}}]
\label{fsxfrloc}
\end{equation}
rather than as in \cite{nagt2}, gauge covariant derivative and
averaging operators could be defined by (\ref{covave}) and
\begin{equation}
(D_\lambda  F^{\mu\nu})_{0000}={1\over h}\left[
(L_\lambda)_1(F^{\mu\nu})_1(L^\dagger_\lambda)_1
-(F^{\mu\nu})_0\right],
\label{covdiff}
\end{equation}
which yields the local  covariant Yang-Mills equation:
\begin{equation}
\prod_{\lambda\ne\nu}\tilde D_\lambda D_\nu F^{\mu\nu}=j^\mu,
\label{ym}
\end{equation}
where we can adopt the following as a gauge-covariant current
(see Sec.~6):
\begin{equation}
(j^\mu)_{\text{m}}=g\overline{\Psi}_{\text{m}}T
\gamma^\mu\Psi_{\text{m}}.
\label{current}
\end{equation}
However, (\ref{fsxfrloc}) {\it does not hold!}  As shown in
Sec.~2 the field strength constructed locally in
terms of the link operators transforms according to (\ref{fsxfr}),
defined at the center of the finite element.  However, given
$F_{\mu\nu}$ constructed in Sec.~2, we can construct a locally
covariant field strength $f_{\mu\nu}$ according to
\begin{equation}
(F_{\mu\nu})_{\overline{\text{m}}}=\left(\prod_\lambda
\tilde D_\lambda f_{\mu\nu}\right)_{\text{m}},
\label{locfs}
\end{equation}
which does transform according to (\ref{fsxfrloc}),
and which then satisfies (\ref{ym}).  Note that, in part,
(\ref{locfs}) undoes the transformation
(\ref{fs}), so that if we choose an appropriate ordering
\begin{equation}
({\cal F}_{\mu\nu})=\tilde D_{\mu}\tilde D_\nu f_{\mu\nu}.
\label{nonlocalf}
\end{equation}
However, as a result of inverting the averaging operators $\tilde
D_\mu$,
that is solving the difference equation (\ref{nonlocalf}),
$f_{\mu\nu}$ is not local, but
depends on vector potentials over the entire prior lattice.  It
appears
impossible to have a completely local formalism.

\section{Axial-Vector Current Anomaly in (1+1) Dimensions}
Let us illustrate the calculational aspects of this scheme in the
simplest context, that of an Abelian theory in which the only
nontrivial element is the Dirac equation (\ref{localdir}).
(We will henceforward replace $g$ by $e$.)
Although the result was stated in \cite{mat}, it is useful to
first revisit the two-dimensional case of the Schwinger model,
with the fermion mass $\mu=0$.  In terms of chiral components,
that is, eigenvectors of $i\gamma_5$ with eigenvalue equal to $\pm1$,
the solution of that equation is (for a particular ordering)
\begin{mathletters}\label{sol}
\begin{eqnarray}
\psi^{(+)}_{m+1,n+1}&=&e^{ieh({\cal A}_0)_{m,n+1}}
e^{ieh({\cal A}_1)_{m+1,n+1}}
\psi^{(+)}_{m,n},\\
\psi^{(-)}_{m,n+1}&=&e^{ieh({\cal A}_0)_{m,n+1}}
e^{-ieh({\cal A}_1)_{m+1,n}}
\psi^{(-)}_{m+1,n}.
\end{eqnarray}
\end{mathletters}
The current is given by (\ref{current}) with $T=1$;
in terms of chiral components
what we wish to compute are the following combinations:
\begin{mathletters}
\begin{eqnarray}
\langle \mbox{``}\partial_\mu j^\mu\mbox{''}
\rangle&=&{1\over h}\langle j^{(+)}_{m+1,n+1}
-j^{(+)}_{m,n}+j^{(-)}_{m,n+1}-j^{(-)}_{m+1,n}\rangle,\\
\langle \mbox{``}\partial_\mu j_5^\mu\mbox{''}\rangle&=&{1\over
h}\langle j^{(+)}_{m+1,n+1}
-j^{(+)}_{m,n}-j^{(-)}_{m,n+1}+j^{(-)}_{m+1,n}\rangle,
\end{eqnarray}
\end{mathletters}
where the quotation marks signify finite-element lattice derivatives.
We use the solution (\ref{sol}) to refer all Dirac fields to the
intermediate
time $n+1$ and we evaluate the fermion matrix elements according
to the Fock space rule\cite{twodqed}
\begin{mathletters}
\label{vevs}
\begin{eqnarray}
\langle\psi^{(\pm)}_{m,n}{}^\dagger\psi^{(\pm)}_{m+q,n}\rangle
&=&\pm q{i\over L}{1\over\sin\pi/M},\qquad \text{for $M$ even},
\\
\langle\psi^{(\pm)}_{m,n}{}^\dagger\psi^{(\pm)}_{m+q,n}\rangle
&=&\pm q{i\over L}{\cos^2\pi/2M\over\sin\pi/M},
\qquad \text{ for $M$ odd},
\end{eqnarray}
\end{mathletters}
where $q=\pm 1$.
In both cases, the vacuum expectation value is taken to zero if
$m=m'$. (As noted in \cite{twodqed} the actual value in that case is
irrelevant.)
Then using (\ref{sol}) in, for example,
\begin{eqnarray}
\left(\prod_\nu \tilde D_\nu\psi^{(+)}\right)_{m,n}&=&
{1\over4}\bigg(\psi^{(+)}_{m,n}
+e^{-ieh({\cal A}_1)_{m+1,n}}\psi^{(+)}_{m+1,n}\nonumber\\
&&\mbox{}+e^{-ieh({\cal A}_0)_{m,n+1}}\psi^{(+)}_{m,n+1}
+e^{-ieh({\cal A}_1)_{m+1,n}}e^{-ieh({\cal A}_0)_{m+1,n+1}}
\psi^{(+)}_{m+1,n+1}\bigg),
\end{eqnarray}
we easily find that the vector current is conserved
exactly\footnote{In \cite{mat} results (\ref{dj}) and
(\ref{dj5}) were established only to $O(h)$.},
\begin{equation}
\langle\mbox{``}\partial_\mu j^\mu\mbox{''}\rangle=0
\label{dj}
\end{equation}
while the axial-vector current is anomalous,
\begin{eqnarray}
\langle\mbox{``}\partial_\mu j_5^\mu\mbox{''}
\rangle&=&{e\over 2M h^2}{1\over\sin\pi/M}
\big[\sin eh({\cal A}_1)_{m+2,n+1}+\sin eh({\cal
A}_1)_{m+1,n+1}\nonumber\\
&&\mbox{}+\sin eh[({\cal A}_0)_{m+1,n+1}-({\cal A}_0)_{m+2,n+1}-
({\cal A}_1)_{m+2,n}]\nonumber\\
&&\mbox{}-\sin eh[({\cal A}_0)_{m+1,n+1}+
({\cal A}_1)_{m+1,n}-({\cal A}_0)_{m,n+1}\big]\nonumber\\
&\approx&{e\over M\sin\pi/M} E,\quad (h\to0),
\label{dj5}
\end{eqnarray}
where $E=\mbox{``}\partial_0A_1-\partial_1 A_0\mbox{''}$ is the
lattice
electric field, the familiar finite-element lattice result
\cite{review}.

The above calculation seems quite similar to that given for the
nonlocal
finite-element formulation in \cite{twodqed}, and is certainly no
simpler.  In fact, the calculations are identical
if, as in \cite{twodqed}, we choose the gauge $A_0=0$.  For then the
massless Dirac equation can be written in terms of the transfer
matrix,
defined by
\begin{equation}
\psi_{n+1}=T\psi_n,
\end{equation}
which is to be understood as a matrix equation in the spatial
coordinates.
Here the transfer matrix is
\begin{equation}
T={1+i\gamma_5{\cal D}\over1-i\gamma_5{\cal D}},
\end{equation}
where the covariant derivative, ${\cal D}=-(h/2)(\tilde
D_1)^{-1}D_1$,
is defined in terms of (the time coordinate $n$ is suppressed)
 \begin{mathletters}
\begin{eqnarray}
(\tilde D_1)_{m,m'}&=&{1\over2}\left(\delta_{m,m'}
+\delta_{m+1,m'}e^{-ieh({\cal A}_1)_{m+1}}\right),
\label{d0}
\\
(D_1)_{m,m'}&=&{1\over h}\left(-\delta_{m,m'}
+\delta_{m+1,m'}e^{-ieh({\cal A}_1)_{m+1}}\right).
\end{eqnarray}
\end{mathletters}
Equation (\ref{d0}) is inverted as in (17) of \cite{twodqed}
with the result that the covariant derivative operator coincides
exactly with that given in (10) of that reference, and hence the
same conclusions follow.

\section{Current Anomalies in (3+1) Dimensions}
We wish to repeat the above calculation in four dimensions.  Again we
will set the mass $\mu=0$, so we have to solve the following symbolic
Dirac equation involving the link operators (\ref{link}) for the
simplest
possible ordering:
\begin{eqnarray}
\gamma^0(L_3+1)(L_2+1)(L_1+1)(L_0-1)\psi&+&\gamma^1
(L_3+1)(L_2+1)(L_1-1)(L_0+1)\psi
\nonumber\\
\mbox{}+\gamma^2(L_3+1)(L_2-1)(L_1+1)(L_0+1)\psi&+&\gamma^3
(L_3-1)(L_2+1)(L_1+1)(L_0+1)\psi=0,
\end{eqnarray}
where, for example, $\psi=\psi_{0000}$,
$L_0\psi=(L_0)_{0001}\psi_{0001}$,
and $L_2L_0\psi=(L_2)_{0100}(L_0)_{0101}\psi_{0101}$.
We must solve this system of equations at each lattice site at a
given time.
For simplicity, let us consider the simplest nontrivial rectangular
spatial lattice, with the number of sites in the 1 direction being
$M_1=2$,
while in the other two directions there is but a single site,
$M_2=M_3=1$.
We anticipate the periodic/antiperiodic boundary condition,
\cite{qed},
\begin{equation}
\psi_{m+M}=(-1)^{M+1}\psi_m,
\label{antibc}
\end{equation}
so we expect that the fields should be antiperiodic in the 1
direction.
The system of Dirac equations  reduces to the following
simple matrix problem
\begin{equation}
R\psi_1=S\psi_0,
\end{equation}
where the Dirac fields at time 1 and time 0 are
\begin{equation}
\psi_1=\left(\begin{array}{c}
\psi_{01}\\
\psi_{11}
\end{array}\right),\quad
\psi_0=\left(\begin{array}{c}
\psi_{00}\\
\psi_{10}
\end{array}\right),
\label{diraccolvec}
\end{equation}
and the matrices are given by
\begin{equation}
\gamma^0R^+=\left(\begin{array}{cccc}
0&-l_0c&2\tilde l_0\tilde l_1 &-\tilde l_0\tilde l_1 c\\
l_0c^*&2l_0&\tilde l_0\tilde l_1c^*&0\\
-2l_0l_1&l_0l_1d&0&-\tilde l_0d\\
-l_0l_1d^*&0&\tilde l_0d^*&2\tilde l_0
\end{array}\right),
\quad \gamma^0R^-=\left(\begin{array}{cccc}
2l_0&l_0c&0&\tilde l_0\tilde l_1 c\\
-l_0c^*&0&-\tilde l_0\tilde l_1c^*&2\tilde l_0\tilde l_1\\
0&-l_0l_1d&2\tilde l_0&\tilde l_0d\\
l_0l_1d^*&-2l_0l_1&-\tilde l_0d^*&0
\end{array}\right),
\label{are}
\end{equation}
and
\begin{equation}
\gamma^0S^+=\left(\begin{array}{cccc}
2&c&0&\tilde l_1 c\\
-c^*&0&-\tilde l_1c^*&2\tilde l_1\\
0&-l_1d&2&d\\
l_1d^*&-2l_1&-d^*&0
\end{array}\right),
\quad\gamma^0S^-=\left(\begin{array}{cccc}
0&-c&2\tilde l_1&-\tilde l_1 c\\
c^*&2&\tilde l_1c^*&0\\
-2l_1&l_1d&0&-d\\
-l_1d^*&0&d^*&2
\end{array}\right),
\label{ess}
\end{equation}
where
\begin{equation}
l_0=(L_0)_{01},\quad \tilde l_0=(L_0)_{11},\quad l_1=(L_1)_{00},\quad
\tilde l_1=(L_1)_{10},
\end{equation}
and
\begin{equation}
c=\tan {eh\over2}({\cal A}_{3})_{00}+i\tan {eh\over2}({\cal
A}_2)_{00},\quad
d=\tan {eh\over2}({\cal A}_{3})_{10}+i\tan {eh\over2}({\cal
A}_2)_{10},
\label{candd}
\end{equation}
and we have denoted eigenvalues of $i\gamma_5$ by the $\pm$
superscripts,
have used the following representation of Dirac matrices:
\begin{equation}
\gamma^0\mbox{\boldmath{$\gamma$}}
=i\gamma_5\mbox{\boldmath{$\sigma$}},
\end{equation}
and  have chosen $\sigma_1$ to be the Pauli $\sigma_z$ matrix.
The transfer matrix $T$ is then given by
\begin{equation}
T=R^{-1}S.
\end{equation}

Unfortunately, it turns out in general the theory is not unitary,
that is $TT^\dagger\ne1$.  This is true even in the temporal gauge,
$A_0=0$.  This is traced back to the failure of the covariant
derivative,
given by (\ref{localdir}),
\begin{equation}
\mbox{\boldmath{${\cal D}$}}={1\over \tilde \Delta_0}
\mbox{\boldmath{$\Delta$}}
\end{equation}
where symbolically
\begin{eqnarray}
\tilde\Delta_0&=&(L_3+1)(L_2+1)(L_1+1),\quad\Delta_1
=(L_3+1)(L_2+1)(L_1-1),
\nonumber\\
\Delta_2&=&(L_3+1)(L_2-1)(L_1+1),\quad\Delta_3=(L_3-1)(L_2+1)(L_1+1),
\end{eqnarray}
to be skew Hermitian.\footnote{It is easily seen that no other
ordering
will resolve this problem.  For the given ordering, ${\cal D}_2=
(L_1+1)^{-1}(L_2-1)^{-1}(L_2+1)(L_1-1)$; the inner factor involving
$L_2$ is clearly skew Hermitian, but the appearance of the $L_1$
terms
destroys that property.}
  In fact, it is easily verified that although
${\cal D}_1$ is skew Hermitian, ${\cal D}_2$ is not unless $l_1\tilde
l_1=1$.
So, to proceed, we will simply choose the gauge with
$A_0=A_1=0$, which should suffice for the anomaly calculation.
This simplifies the transfer matrix dramatically by replacing
\begin{equation}
l_0\to1,\quad \tilde l_0\to 1,\quad l_1\to1,\quad \tilde l_1\to1.
\label{linksub}
\end{equation}
 Then it is easily seen that $T$ is unitary.
(If $\psi$ were periodic rather than antiperiodic in the 1 direction,
which would be accomplished by changing the sign of $l_1$ in
(\ref{are}) and (\ref{ess}), $T$ would not be unitary.)
In fact, it will suffice in the following to expand $T$ to bilinears
in
$c$ and $d$.  It then turns out to be
\begin{equation}
T^+\approx\left(\begin{array}{cccc}
\upsilon&d&-1+\xi&0\\
-c^*&\upsilon&0&1+\tilde\xi\\
1+\tilde\xi&0&\upsilon^*&c\\
0&-1+\xi&-d^*&\upsilon^*
\end{array}
\right),\quad T^-\approx\left(\begin{array}{cccc}
\upsilon^*&-c&1+\tilde\xi&0\\
d^*&\upsilon^*&0&-1+\xi\\
-1+\xi&0&\upsilon&-d\\
0&1+\tilde\xi&c^*&\upsilon
\end{array}\right),
\end{equation}
where we have abbreviated
\begin{equation}
\xi={1\over2}dd^*,\quad
\tilde\xi=-{1\over2}cc^*,\quad
\upsilon=-{1\over2}dc^*.
\label{defs}
\end{equation}
In the following we will use the block form of $T$, which refers to
specific spatial coordinates:
\begin{equation}
T=\left(\begin{array}{cc}
T_{00}&T_{01}\\
T_{10}&T_{11}
\end{array}\right).
\label{blockform}
\end{equation}

We now take matrix elements of the current in Fock-space states
defined
in terms of the momentum-space expansion of the free Dirac field
\cite{twodqed,mmsb}
\begin{equation}
\psi_{m,1}=\sum_{s,p}\sqrt{{\mu\over\omega}}
\left(b_{ps}u_{ps}e^{i(p+1/2)
m2\pi/M}+d^\dagger_{ps}v_{ps}e^{-i(p+1/2)m2\pi/M}\right).
\end{equation}
Because of the particular choice of gauge (\ref{linksub}) the current
is proportional to
 the usual finite-element one, and the matrix element of $j^0$,
for example, is
\begin{equation}
\langle j^0_{m,n}\rangle=C{e\over16}\langle[(\psi^\dagger_1(1+T))_1+
(\psi^\dagger_1(1+T))_0][((1+T^\dagger)\psi_1)_1
+((1+T^\dagger)\psi_1)_0]\rangle,
\end{equation}
where the Dirac fields are those given in (\ref{diraccolvec}),
\begin{equation}
C=\cos^2{eh\over2}({\cal A}_2)_{00}\cos^2{eh\over2}({\cal A}_3)_{00},
\end{equation}
 and the matrix subscripts refer to the spatial coordinate.
The matrix elements are evaluated using the following easily derived
formula for $M=2$:
\begin{equation}
\langle\psi^\dagger_m\Gamma\psi_{m'}\rangle={1\over h^3}(\delta_{mm'}
\mbox{tr}\, \Gamma+2i\epsilon_{mm'}\mbox{tr}\, \gamma^0\gamma^1
\Gamma).
\end{equation}
Then a straightforward calculation gives the result
\begin{equation}
\langle j^0_{m,n}\rangle={e\over h^3}(1+\lambda)C,
\end{equation}
where
\begin{equation}
\lambda={1\over4}(dd^*-cd^*-dc^*-cc^*).
\label{lambda}
\end{equation}
 A similar calculation reveals that $\langle j^1\rangle=0$.
The axial-vector current matrix elements vanish.  Consequently, for
this tiny lattice, there is no axial-vector current anomaly, but
there is a vector current anomaly,
\begin{eqnarray}
\langle \mbox{``}\partial_\mu j^\mu\mbox{''}\rangle&=&
-{e^3\over8h^2}\{[({\cal A}_2)_{11}^2+
({\cal A}_2)_{11}({\cal A}_2)_{01}+({\cal A}_2)_{01}^2
-({\cal A}_2)_{10}^2-({\cal A}_2)_{10}({\cal A}_2)_{00}-({\cal
A}_2)_{00}^2]
\nonumber\\
&&\quad+[2\leftrightarrow3]\},
\label{matresult}
\end{eqnarray}
a somewhat curious result which will be discussed below.

\section{Finite-Element Current Anomaly Calculation}
Because of the various difficulties seen in the calculation exhibited
in the previous section, we return now to original formulation.
There, the {\it unitary\/} transfer matrix in (3+1) dimensions is, in
the gauge
$A_0=0$, \cite{mmsb}
\begin{equation}
T={1+\gamma^0\mbox{\boldmath{$\gamma$}}\cdot{\cal D}\over
1-\gamma^0\mbox{\boldmath{$\gamma$}}\cdot{\cal D}},
\label{fetee}
\end{equation}
where (the time coordinate $n$ is suppressed)
\begin{equation}
{\cal D}^i_{m_i,{\bf m}_\perp;m'_i,{\bf m}'_\perp}
=-(-1)^{m_i+m_i'}(\epsilon_{m_im_i'}\cos\hat\zeta_{m_im_i'}-i
\sin\hat\zeta_{m_im_i'})\sec\zeta^{(i)}\delta_{{\bf m_\perp
m'_\perp}},
\label{covariantderiv}
\end{equation}
where (and now all the local dependence on the other spatial
coordinates
is also suppressed)
\begin{equation}
\zeta_{m_i}={eh\over2}{\cal
A}^i_{m_i},\quad\zeta^{(i)}=\sum_{m_i=1}^M
\zeta_{m_i},
\end{equation}
and
\begin{equation}
\hat\zeta_{m_i,m_i'}=\sum_{m''_i=1}^M\mbox{sgn}\,(m''_i-m_i)\,
\mbox{sgn}\,(m''_i-m'_i)\zeta_{m''_i}.
\end{equation}
Here,  the sign function is
\begin{equation}
\mbox{sgn}(m)=\left\{\begin{array}{cc}
1,&m>0,\\
-1,&m\le0.
\end{array}\right.
\end{equation}
Again, let us consider the smallest possible nontrivial lattice,
with the number of lattice points in the 1 direction being $M_1=2$,
while the 2 and 3 directions have but one site, $M_2=M_3=1$.
This leads again to an $4\times4$ transfer matrix for each chirality,
which for small
$ehA$ is, in the block form given in (\ref{blockform}),
\begin{mathletters}
\label{tfe}
\begin{eqnarray}
T^\pm_{00}&=&\left(\begin{array}{cc}
\lambda&\pm{c+d\over2}\\
\mp{c^*+d^*\over2}&\lambda
\end{array}\right),\quad
T_{01}^+=T_{10}^{-\dagger}={s\over2}\left(\begin{array}{cc}
\alpha&\beta\\
\beta^*&-\alpha^*
\end{array}\right),\\
T_{10}^+=T_{01}^{-\dagger}&=&{s^*\over2}\left(\begin{array}{cc}
\tilde\alpha&\tilde\beta\\
\tilde\beta^*&-\tilde\alpha^*
\end{array}\right),\quad
T_{11}^\pm=\left(\begin{array}{cc}
-\tilde\lambda&\pm{c+d\over2}\\
\mp{c^*+d^*\over2}&-\tilde\lambda
\end{array}\right).
\end{eqnarray}
\end{mathletters}
Here $\lambda$ is defined above in (\ref{lambda}),
\begin{equation}
\tilde\lambda={1\over4}(dd^*+cd^*+dc^*-cc^*),
\end{equation}
$c$ and $d$ are as in (\ref{candd}),
\begin{mathletters}
\label{transfer}
\begin{eqnarray}
\alpha&=&{1\over2}(4+4it_1-4t_1-dd^*+d^*c-c^*d-cc^*),\\
\tilde\alpha&=&{1\over2}(-4-4it_1+4t_1+dd^*+d^*c-c^*d+cc^*),\\
\beta&=&=d(1+it_1)-c(1-it_1),\\
\tilde\beta&=&=d(1-it_1)-c(1+it_1),
\end{eqnarray}
\end{mathletters}
and
\begin{equation}
t_i=\tan\zeta^{(i)},\quad
s=-e^{i(\zeta^1_1-\zeta^1_2)}\sec\zeta^{(1)}.
\end{equation}
Even though the transfer matrices are quite different
(the form in (\ref{tfe}) is more complicated because $A_1\ne0$),
the results of
the calculation are very similar: The axial-vector anomaly is zero,
while the vector anomaly is
\begin{equation}
\langle\mbox{``}\partial_\mu j^\mu\mbox{''}\rangle=
-{e\over2h^4}[(t_2)_{21}(t_2)_{11}+(t_3)_{21}(t_3)_{11}
-(t_2)_{20}(t_2)_{10}-(t_3)_{20}(t_3)_{10}].
\label{feresult}
\end{equation}
Here, the current is not (\ref{current}) but
\begin{equation}
j^\mu_{{\bf m},n}=e\overline\psi_{\overline{\bf m},\overline{n}}
\gamma^\mu\psi_{\overline{\bf m},\overline{n}},
\label{fecurrent}
\end{equation}
the averaging being over the finite element without the link
operators.
The result (\ref{feresult}), which  is manifestly gauge invariant,
appears to be, like (\ref{matresult}), a lattice
artifact.\footnote{It is
probable that this result reflects the rectangular nature of the
lattice
considered here.  A similar ``anomalous'' anomaly was found when the
space-time lattice was not chosen to be square: That is, when the
lattice
spacing in the time direction was not equal to that in the space
directions.
See \cite{twodqed}.}
It is not a lattice version of $F^2$, as
one might anticipate.  However, we should not be discouraged, since
the
calculation given is for a truly tiny, unrealistic lattice.
 The extension of this calculation to larger
lattices will be presented elsewhere.

\section{Hermiticity, Unitarity, and the Existence of a Symmetry
Current}

In this paper we pursued a variation on the finite-element
formulation
of lattice gauge theory which seemed at first sight very promising.
The idea was to use the link variables, in term of which the local
field strength was constructed, to express the Dirac and Yang-Mills
equations in local form, rather than the form involving the entire
prior lattice given previously.  A locally gauge-invariant
construction
can indeed be done, one which is inequivalent to the earlier
formulation.
However, the new scheme is less advantageous than it first seemed,
and
ultimately fails to be consistent:
\begin{itemize}
\item The field strength which appears in the Yang-Mills equation
is not locally constructed in terms of local link operators.  It
appears impossible to have a local formulation consistent with local
non-Abelian gauge symmetry.
\item Moreover, detailed calculations, even in the Abelian theory
where the equations are local, turn out to be no simpler, and perhaps
more complicated than those in the original formulation.
\item Disastrously, unitarity is violated.  Explicitly, we have
seen in the Abelian case that the transfer matrix is not unitary,
even in a temporal gauge.  This is traced to the fact that the
covariant
derivative operators are not skew Hermitian.  In contrast, the
original
formulation is manifestly unitary (canonical).  It should be recalled
that
preservation of the canonical commutation relations at the lattice
sites
was the original motivation for adopting the finite-element
prescription
for field theory on a Minkowski lattice.
\end{itemize}

We can interpret these results in a positive light.  It was always
apparent
that, although some arbitrary choices had to be made to implement
local
gauge invariance, the requirement of that invariance, that is, that
the
transformation equations could be ``integrated,'' was rather rigid.
The
findings presented here strengthen that conclusion.  The fact that we
have now shown that a rather natural and attractive alternative
gauging process is ultimately inconsistent makes the pursuit of
extracting
physical information from the orginal, consistent, approach that much
more compelling.

Finally, we should add some remarks about the current employed in
both formulations studied here.  It should be noted that the choice
of current was essentially arbitrary, subject only to the requirement
that it be locally gauge covariant.  It is essential to note that
{\it the current cannot be derived from the Dirac equation.}  In
fact,
because our equations of motion cannot be derived from an action,
there is no connection between symmetry (say chiral symmetry) and
conservation laws (say axial-vector current conservation).
In our consistent formulation this is because the Dirac equation
is asymmetric between past and future.  If one were interested in
a Euclidean formulation, one would choose a Dirac equation symmetric
in the fourth coordinate, and an action, and corresponding current
could be constructed.  That current would, however, also  involve
all lattice sites in the fourth coordinate, and therefore would be
be unusable in the Minkowski context, where we wish to solve the
operator equations of motion by time-stepping through the lattice.
This Euclidean construction and its Minkowski failure is sketched in
the Appendix.

\section*{Acknowledgments}
I thank the UK PPARC for the award of a Senior Visiting
Fellowship and Imperial College for its hospitality during
the initial stages of writing this paper.
I further thank the US Department of Energy and the University
of Oklahoma College of Arts and Sciences for partial
financial support of this research.  I am grateful to Maarten
Golterman, Simon Hands, and Tai Wu for useful conversations.

\appendix
\section*{Current constructed from Euclidean Lagrangian}
In the text we simply assumed a form of the current (\ref{current})
and (\ref{fecurrent})
which was manifestly gauge invariant.  We are at liberty to do
so, because the Minkowski finite-element equations of motion are
not derivable from a Lagrangian.  The current cannot be derived
from the Dirac equation, but is an independent source for the Maxwell
equations, see Ref.~\cite{qed,nagt}.  However, if one were to work in
Euclidean space-time (periodic  or antiperiodic in all four
directions),
it is possible to construct an action
from which the equations of motion are derivable, and which therefore
supplies a lattice current.  The fermion part of that action is
(a factor of $i$ is absorbed in going to Euclidean space)
\begin{equation}
W_f=h^4\sum_{\bf m, m'}\psi^\dagger_{\overline{\bf m}}
\left({2\over h}\bbox{\Gamma\cdot{\cal D}}
+i\mu\right)_{\bf m,m'}\psi^{\vphantom\dagger}
_{\overline{\bf m}'},\quad
\bbox{\Gamma}=(\gamma^0\gamma_k,\gamma^0).
\label{action}
\end{equation}
If we vary (\ref{action}) with respect to $\psi^\dagger$ we
obtain the Euclidean Dirac equation
\begin{equation}
\left({2\over h}\bbox{\Gamma\cdot{\cal D}}
+i\mu\right)\psi=0.
\label{euclideq}
\end{equation}
where, as in (\ref{action}), a four-dimensional scalar product is
implied.

Given an action, we can construct a conserved vector current
by making a local gauge transformation
\begin{equation}
\delta\psi_{\overline{\bf m}}=
ie\delta\Omega_{\bf m}\psi_{\overline{\bf m}}.
\label{gt}
\end{equation}
Because the Dirac equation, and hence the action, is invariant
under the global version of (\ref{gt}), $\delta\Omega_{\bf m}
=\delta\Omega= \text{constant}$,  we must have by the action
principle
\begin{equation}
\delta W_f=0=-h^4\sum_{\bf m}J^i_{\bf m}
{1\over h}(\delta\Omega_{m_i,{\bf m}_\perp}
-\delta\Omega_{m_i-1,{\bf m}_\perp}),
\end{equation}
from which we read off the  conserved current
\begin{eqnarray}
J^i_{\bf m}=-e&&\sum_{m_i',m_i''}
\psi_{\overline{m}_i',
\overline{\bf m}_\perp}^\dagger \Gamma^i
\psi_{\overline{m}''_i,\overline{\bf m}_\perp}^{\vphantom\dagger}
\,\text{sgn}(m_i-m_i')\,\text{sgn}(m_i-m_i'')\nonumber\\
&&\times(-1)^{m'_i+m_i''}\sec\zeta^{(i)}\exp(-i\epsilon_{m_i',m_i''}
\hat\zeta_{m_i',m_i''}).
\label{eculidcur}
\end{eqnarray}
(The same result, of course, can be obtained by  varying
${\cal D}$ (\ref{covariantderiv})
 with respect to ${\cal A}^i_{{m_i},{\bf m}_\perp}$.)
The expression for this Euclidean current has been simplified
by deleting constant terms.  It is easy to verify explicitly
that this current is both conserved and gauge invariant.
Similarly, by making a chiral transformation,
\begin{equation}
\delta\psi_{\overline{\bf m}}=\gamma^0\gamma_5\delta\Omega_{\bf m}
\psi_{\overline{\bf m}},
\end{equation}
we can construct the axial-vector current $J^i_{5{\bf m}}$, which
has the form of (\ref{eculidcur}) with the replacement
\begin{equation}
e\Gamma^i\to \gamma^0i\gamma_5\Gamma^i\equiv \Gamma_5^i,\quad
\Gamma_5^i=(-i\gamma_5\gamma^i, -i\gamma_5).
\end{equation}
By construction, these currents possess no anomalies.  However,
they appear to be completely unacceptable, because they are
horribly nonlocal.  In particular, they possess no Minkowski
analogues,
in the sense that it is not possible to analytically continue
back to real unbounded times.
Crucial to our formulation is the propagation of the operators
from past times to the present time, so that we can solve for the
field operators by time-stepping through the lattice.
The Euclidean current
(\ref{eculidcur}) involves fermion field operators at all
Euclidean times, which  would make it impossible to solve
for the operators at time $n$ in terms of operators at earlier
times.  Therefore, for the considerations of the
text we use the gauge-invariant current (\ref{current})
and (\ref{fecurrent}) and their
axial analogues, currents which can and do possess anomalies.

It is further illuminating to note that if we were to use the current
(\ref{eculidcur}) in a one-loop lattice calculation of the vacuum
polarization
in two dimensions, we would find a vanishing anomaly, rather than the
value $e^2/\pi$ reported in \cite{twodqed}.  This is because species
doublers occur in the action defined by (\ref{action}).  Such
doublers
are absent in the finite-element scheme based on equations of motion
and the current (\ref{fecurrent}).

\end{document}